# 1SWASP J022916.91-395901.4 – a Possible New VY Sculptoris Variable in Eridanus

STEFAN HÜMMERICH [1,4,5], KLAUS BERNHARD [2,4,5], GREGOR SRDOC [3]

1) D-56338 Braubach, Germany; e-mail: ernham@rz-online . de
2) A-4030 Linz, Austria; e-mail: klaus . bernhard@liwest . at
3) 51216 Viškovo, Croatia; e-mail: gregorsrdoc@gmail . com
4) Bundesdeutsche Arbeitsgemeinschaft für Veränderliche Sterne e.V. (BAV), Munsterdamm 90, D-12169 Berlin, Germany
5) American Association of Variable Star Observers (AAVSO), 49 Bay State Road, Cambridge, MA 02138, USA

BAV Mitteilungen Nr. 235

**Abstract:** We report the discovery of 1SWASP J022916.91-395901.4 = GSC 07552-00389, a possible new VY Sculptoris variable in Eridanus, which is associated with the X-ray source 1RXS J022917.1-395851.

## 1. Introduction

Cataclysmic variables (CVs) are close binary systems that typically consist of a white dwarf accreting matter from a low-mass secondary companion that fills its Roche lobe (e.g. Warner 1995; Leach et al. 1999). Among the several types of CVs, the VY Sculptoris stars – which are usually listed under the nova-like variables (GCVS-type NL) – do not show nova-like outbursts but are notorious for exhibiting significant fading events from an otherwise approximately constant maximum magnitude. Hence, they are sometimes referred to as "anti-dwarf novae" (e.g. Leach et al. 1999). These reductions in brightness are thought to arise from a temporary reduction or cessation of mass transfer (e.g. Warner 1995; Sion et al. 2009).

During an investigation of optical counterparts of X-ray sources from the ROSAT catalogues (ROSAT All-Sky Bright Source Catalogue, Voges et al. 1999; ROSAT All-Sky Survey Faint Source Catalogue, Voges et al. 2000), we have identified the star 1SWASP J022916.91-395901.4 = GSC 07552-00389 as a likely VY Scl variable. An overview providing essential data of the object is given in Table 1. The full range of available data for 1SWASP J022916.91-395901.4 is presented in Section 2 and discussed in detail in Section 3. We conclude in Section 4.

| Identifiers | 1SWASP J022916.91-395901.4 |
|---|---|
| | ASAS J022916-3959.0 |
| | SSS_J022917.0-395901 |
| | GSC 07552-00389 |
| | 1RXS J022917.1-395851 |
| Right Ascension (J2000) | 02:29:16.945 |
| Declination (J2000) | -39:59:01.65 |
| Magnitude Range (V) | 12.8 - 16.7: |
| Periods | 0.136621 ±0.00005 d |
| | 0.137206 ±0.00005 d |
| | ~26 d |
| Colour indices | $(B-V)_{APASS}$ = 0.06 |
| | $(B-V)_{SPM4.0}$ = 0.02 |
| | $(J-K)_{2MASS}$ = 0.72 |
| Remarks | Significant fading around HJD 2455000. |

Table 1 - Essential data of 1SWASP J022916.91-395901.4. Positional data were taken from UCAC4 (Zacharias et al. 2012). Colour indices were derived from APASS (Henden et al. 2012), SPM4.0 (Girard et al. 2011) and 2MASS (Skrutskie et al. 2006).

## 2. Data

1SWASP J022916.91-395901.4 = GSC 07552-00389 has been found to exhibit a significant fading event between HJD 2454700 and HJD 2455200, dropping sharply by about 3.5 mag (V∗)[1], which has been well documented by observations from the Siding Spring Survey (SSS; Drake et al. 2009). The long-term light curve of the star, based on all available data, is shown in Figure 1.

The star also exhibits variability on shorter time-scales. This becomes especially obvious in data from the SuperWASP project (SWASP; Butters et al. 2010) which boast very good time sampling. In order to increase the accuracy of the SWASP measurements, the data were binned (bin-size: 0.01 day) and obvious outliers were removed by a visual inspection. SWASP data are made up of two parts, which are separated by an observational gap of ~240 days. The long-term SWASP light curve is shown in Figure 2; properties of the SWASP dataset are given in Table 2.

---

[1] Magnitudes derived from the Siding Spring Survey (Drake et al. 2009) and the SuperWASP project (Butters et al. 2010) are unfiltered values that have been calibrated against V magnitudes and are commonly listed amongst passbands as "CV" (cf. e.g. http://www.aavso.org/vsx/index.php?view=about.passbands). In order to avoid confusion with the abbreviation for cataclysmic variables throughout this paper, we have refrained from using this convention and have chosen to denote the corresponding magnitudes as "V∗" instead.

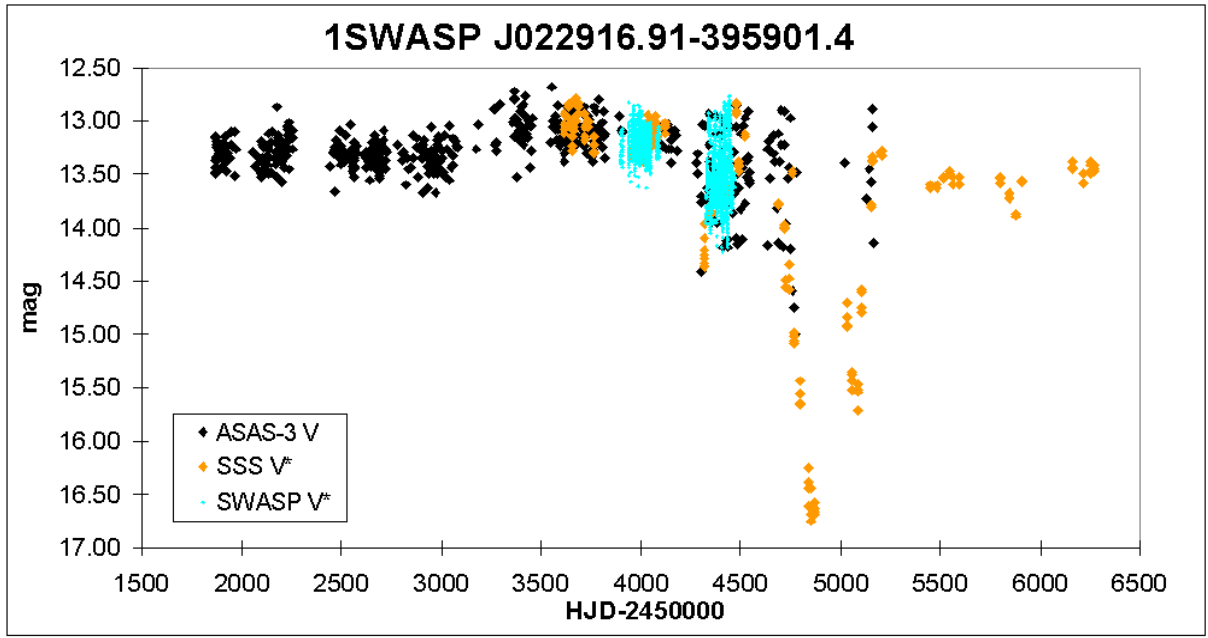

Figure 1 – Light curve of 1SWASP J022916.91-395901.4, based on data from various sky surveys, as indicated in the legend (inset).

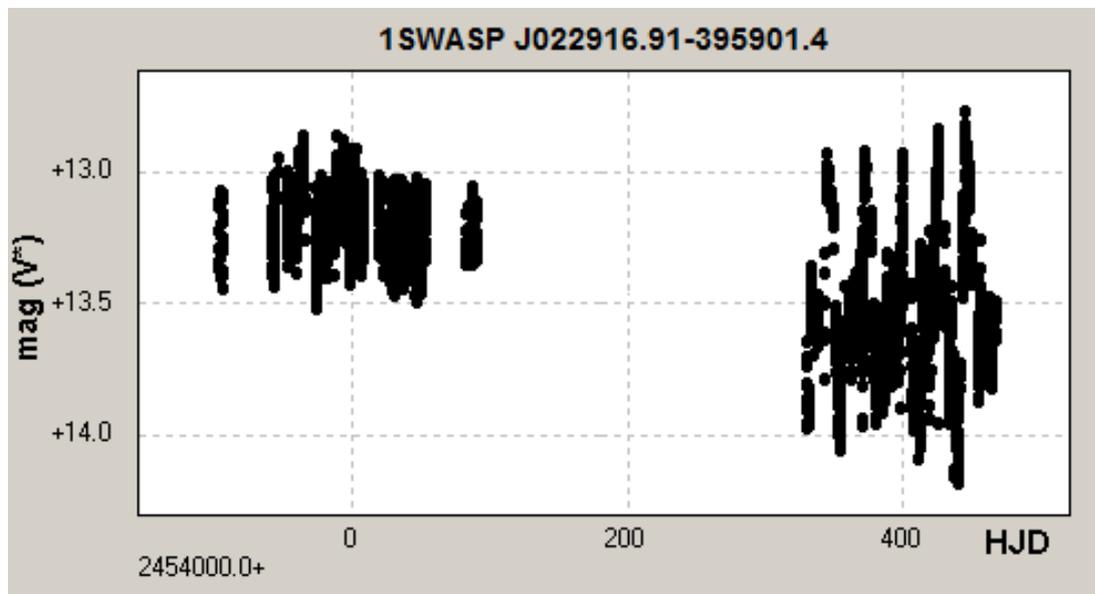

Figure 2 – Long-term light curve of 1SWASP J022916.91-395901.4, based on SWASP data (binned; bin-size: 0.01 days). Obvious outliers have been removed by a visual analysis.

| SWASP dataset | part 1 | part 2 |
| --- | --- | --- |
| time (HJD) | 2453903.628 to 2454090.409 | 2454330.463 to 2454466.451 |
| covered timespan (d) | ~187 | ~136 |
| number of datapoints | 1184 (after binning) | 1756 (after binning) |
| mean magnitude (V*) | 13.20 | 13.53 |
| amplitude (V*) | ~0.8 mag | ~1.4 mag |
| dominant periods (d) | 0.136621 ±0.00005 | 0.137206 ±0.00005<br>~26 |

Table 2 – Properties of the SWASP dataset.

Both parts of SWASP data are distinctly different. Obviously, a mean magnitude shift and an increase in amplitude took place during the observational gap (cf. Figure 2, Table 2), which is reflected in data from other sky surveys. Similar phenomena are also documented during other parts of the long-term light curve of the star that have not been covered by SWASP (cf. Figure 1). Thus, we conclude that the observed differences between both parts of the dataset are intrinsic to the star and not of instrumental origin. Furthermore, we have checked for possible light contamination by another variable star in the same field, which might have affected measurements from the sky surveys. However, no nearby stars of sufficient brightness are present that might induce a significant light contamination.

The more recent SWASP data are characterized by obvious semi-regular brightenings with an amplitude of roughly 1 magnitude ($V*$) and a period around 26 days. Corresponding peaks are also found in the power spectra of ASAS-3 and SSS data, albeit of considerably weaker power. We have searched for additional frequencies using Period04 (Lenz and Breger 2005) and the Lomb-Scargle method in PERANSO (Vanmunster 2007), which resulted in the discovery of regular variations with an amplitude of about 0.2 mag ($V*$) and a period of $P1 = 0.136621 \pm 0.00005$ days[2] in the first part of SWASP data. This signal is still present in the second part of the data, although it is now dominated by another signal at $P2 = 0.137206 \pm 0.00005$ days (cf. Table 2; Figure 3). After excluding the latter oscillation from the data via pre-whitening, P1 once again resumes its dominance.

Phase plots illustrating the observed variability pattern of 1SWASP J022916.91-395901.4 in SWASP data are given in Figure 4. Figure 5 shows details of the SuperWASP light curve which illustrate the short-term variability; here, small cycle-to-cycle variations are visible which are probably the result of the beating of the two detected periods.

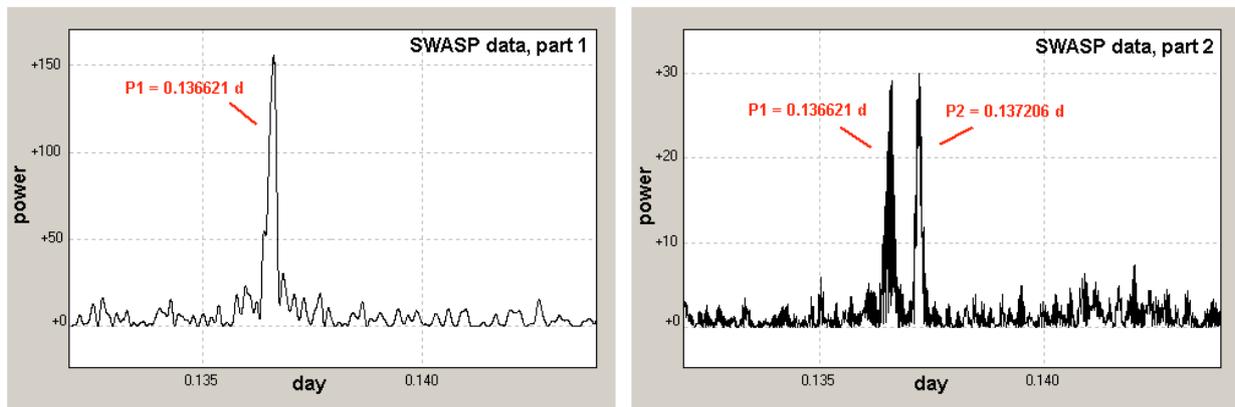

Figure 3 – Lomb-Scargle power spectra for the SWASP data set. The left panel illustrates the power spectrum of the first part of SWASP data (2453900 < HJD < 2454090); the right panel shows the power spectrum of the second part (2454330 < HJD < 2454470). The first part is dominated by a single peak at P1 = 0.136621 days. This signal is still present in the second part, although it is now dominated by another signal at P2 = 0.137206 d.

---

[2] Period uncertainties were estimated using Period04 and employing the width of the corresponding peaks in the power spectra.

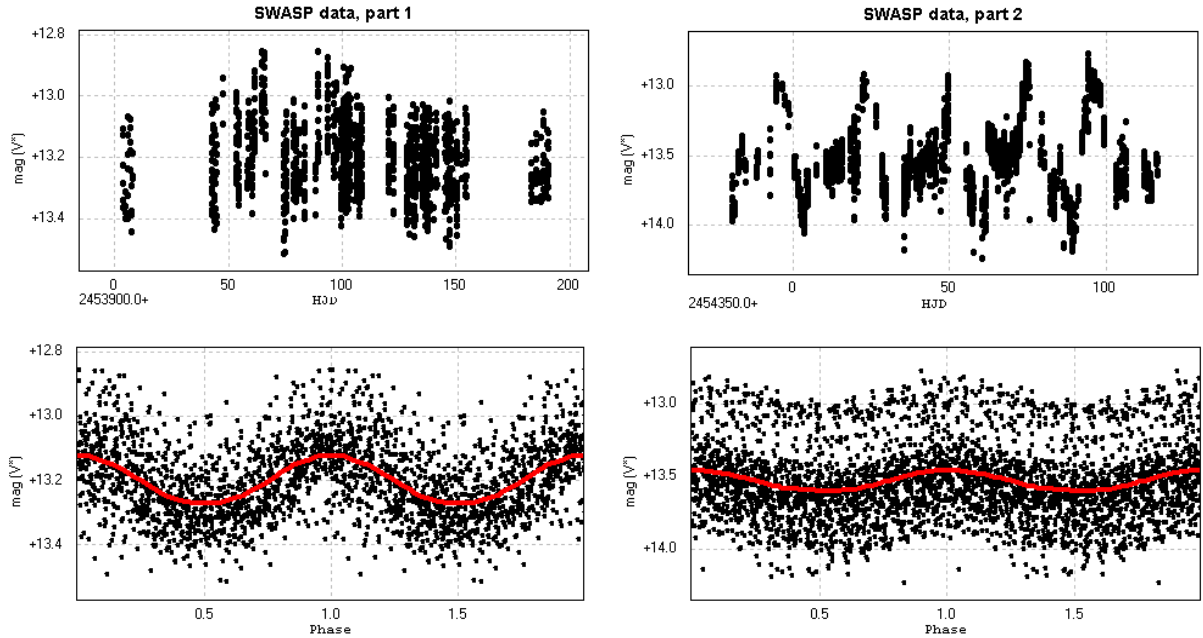

Figure 4 – Short-term variability of 1SWASP J022916.91-395901.4. The left panels show the first part of SWASP data (top; 2453900 < HJD < 2454090) and the same data folded with a period of P1 = 0.136621 days (bottom). The right panels show the second part of SWASP data (top; 2454330 < HJD < 2454470) and the same data folded with a period of P2 = 0.137206 days (bottom). SWASP data have been binned (bin-size: 0.01 days) and obvious outliers have been removed by a visual inspection; fit curves are shown in red. Note the modulation of the light curve with a period of ~26 days during the second part of SWASP coverage.

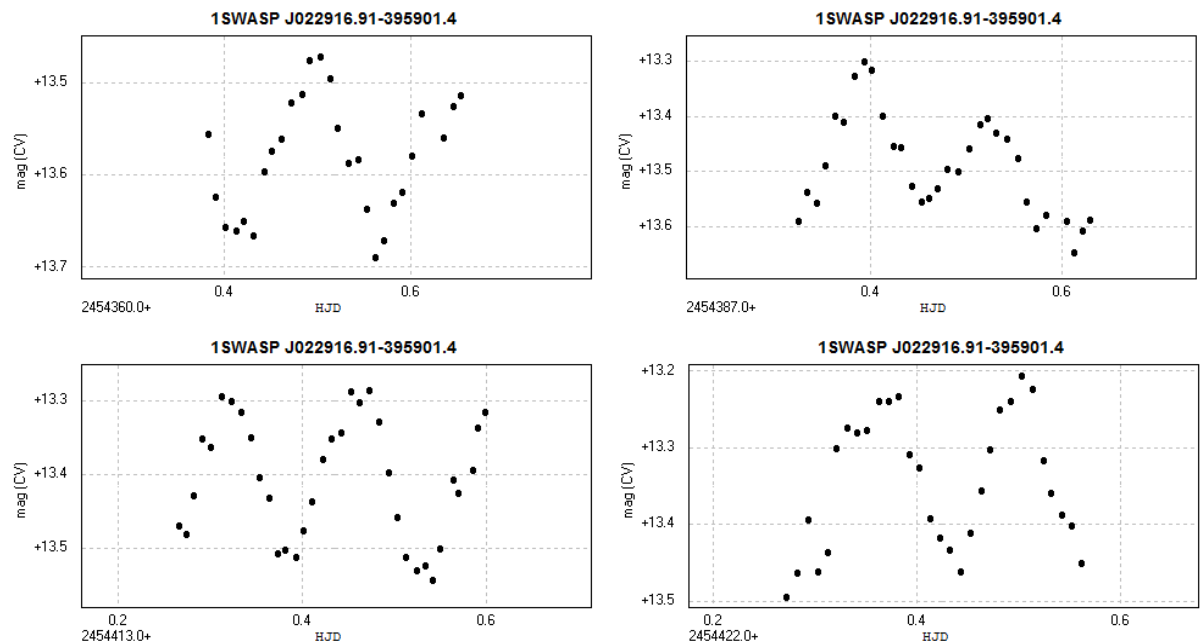

Figure 5 – Light curve details of 1SWASP J022916.91-395901.4, illustrating the short-term variability during the second part of SWASP coverage (2454330 < HJD < 2454470). The diagram has been based on SuperWASP data (binned, bin-size: 0.01 days; obvious outliers have been removed by a visual inspection).

## 3. Discussion

The observed colour indices for 1SWASP J022916.91-395901.4 are rather discrepant. The star is quite blue in APASS and SPM4.0 data – (B-V) = 0.06 and (B-V) = 0.02, respectively – while the observed (J-K) index of 0.72 (2MASS) points to a later spectral type and might be indicative of infrared excess.

The star is even bluer in YB6 data (accessed through the NOMAD catalogue; Zacharias et al. 2005), which indicate (B-V) ≈ -0.4. YB6 measurements are photographic magnitudes based on the scans of blue and yellow plates obtained during the Lick and Yale Proper Motion surveys. As the corresponding plates were often not taken at the same epoch, further uncertainties are added to those deriving from the plate reduction processes (Otero 2014; private communication). In order to investigate this issue, we have compared (contemporaneous) CCD measurements from APASS with measurements derived from the YB6 catalogue via NOMAD for several VY Scl stars. Interestingly, (B-V) indices derived from YB6 are on the average ~0.3 mag bluer than those derived from APASS measurements. Although this conclusion obviously suffers from small number statistics, we are inclined to attribute the difference in colour to the aforementioned sources and feel justified in preferring the CCD measurements of APASS and SPM4.0.

Based on the calculations of Schlafly and Finkbeiner (2011), we estimate an interstellar extinction of $A_V \approx 0.06$ mag; thus, line-of-sight extinction to 1SWASP J022916.91-395901.4 is negligible and should not interfere significantly with the observed colour indices. It is interesting to note that the V measurements from ASAS-3 and the magnitudes from the SSS and the SWASP project, which are unfiltered values that have been calibrated against V magnitudes, match perfectly (cf. Figure 1). Thus, no attempt has been made to shift SWASP and SSS data to the V scale.

1SWASP J022916.91-395901.4 is coincident with the X-ray source 1RXS J022917.1-395851 from the ROSAT All-Sky Survey Faint Source Catalog (Voges et al. 2000) and has also been detected by the ultraviolet (UV) space telescope Galaxy Evolution Explorer (GALEX; Bianchi et al. 2011; FUV = 13.39 mag; NUV = 13.24 mag).

The observed parameters, notably colour indices, X-ray luminosity and variability pattern, provide strong evidence against a classification of 1SWASP J022916.91-395901.4 as a variable star of UX Orionis or R Coronae Borealis type – other groups of variables notorious for exhibiting significant fading events. On the other hand, all observed parameters are in agreement with a classification as a nova-like variable of VY Sculptoris type, which will be enlarged on in the following paragraphs.

The X-ray and UV detection of 1SWASP J022916.91-395901.4 are indicative of high-energy phenomena taking place on the star. In the hardness ratio (HR) diagram, which plots HR1 versus HR2, the star is situated well inside the realm of the CVs and the observed values are in excellent agreement with other VY Sculptoris variables (HR1= 0.77±0.31; HR2= 0.19±0.28; cf. Figure 6).

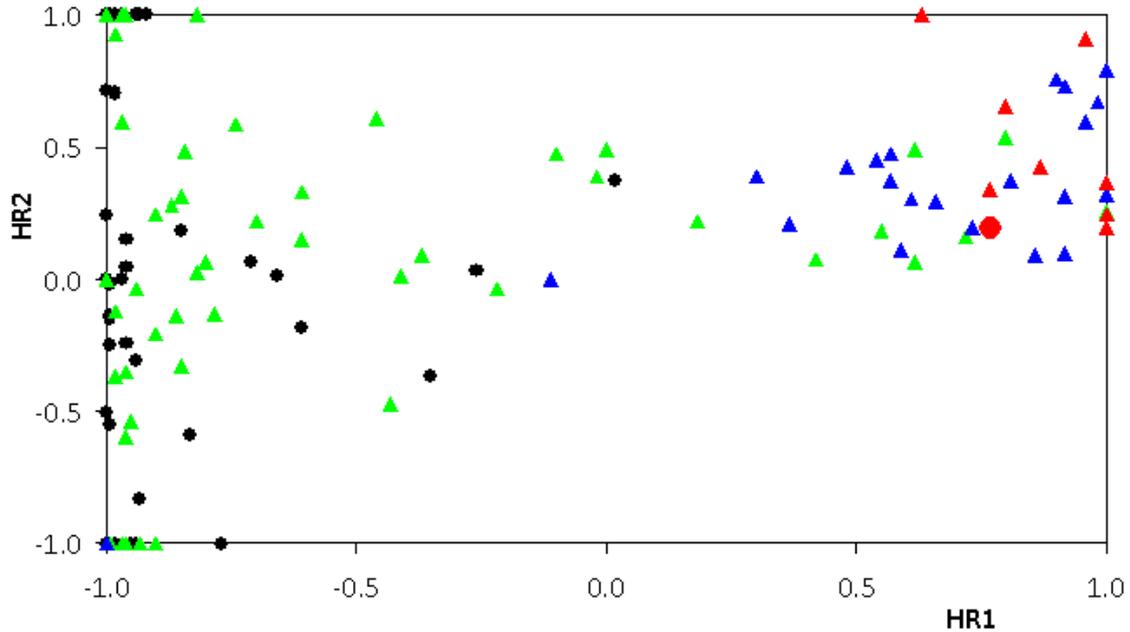

Figure 6 – HR1 versus HR2 diagram, based on data from the ROSAT catalogues (Voges et al. 1999; Voges et al. 2000). The diagram shows the position in hardness ratio space for (non-interacting) white dwarfs (black circles; data from Fleming et al. 1996), AM Herculis systems (green triangles), U Geminorum variables (blue triangles) and VY Sculptoris variables (red triangles). The position of 1SWASP J022916.91-395901.4 is marked by the big red dot. CV classifications were taken from the AAVSO International Variable Star Index (VSX; Watson et al. 2006).

VY Scl systems are generally thought to comprise a hot white dwarf accumulating matter via an accretion disk (Warner 1995). This scenario is in general accord with the observed colour indices. Furthermore, according to Honeycutt and Kafka (2004), the average (B-V) index for nova-like CVs is ~0.00, which holds true for most VY Scl systems like e.g. TT Ari (0.00), BB Dor (0.05), KR Aur (-0.07), BZ Cam (0.03), V794 Aql (0.06) etc. This is in good agreement with the corresponding value for 1SWASP J022916.91-395901.4.

The period of the sinusoidal short-period variability ($P \approx 0.137$ days) observed in 1SWASP J022916.91-395901.4 is in very good agreement with the orbital periods observed in VY Sculptoris systems which show a strong correlation and seem to exclusively range between ~0.133 and ~0.166 days (Warner 1995; Hameury and Lasota 2002; Honeycutt and Kafka 2004). No eclipse features seem to be present in the phase plot (cf. Figure 4); in fact, shape and amplitude of the observed variability are reminiscent of the findings for non-eclipsing VY Scl systems, like e.g. IM Eri (e.g. Armstrong et al. 2013), which often exhibit a single, sinusoidal wave in the range of the orbital period. The resulting photometric variability may be further affected by the superhump phenomenon, which is commented on below. It seems likely, then, that 1SWASP J022916.91-395901.4 is a non-eclipsing system, too, although radial velocity studies are needed to settle this question with authority.

Cataclysmic variables of short orbital period often exhibit photometric variability with a period close to, but not quite matching, the orbital period. This variability is generally interpreted as being due to so-called superhumps (cf. e.g. Warner 1995). Generally, superhump periods slightly exceed orbital periods (e.g. Skillman et al. 1998), although the opposite has been reported as well (e.g. Kozhevnikov 2007). Some bright nova-like variables routinely exhibit superhumps during their normal brightness state (Kozhevnikov 2007), and it is possible that the observed short-period variability of 1SWASP J022916.91-395901.4 might be due to this phenomenon.

Abrupt period changes due to superhumps have been observed in many VY Scl stars, e.g. TT Ari, and are thought to arise from the development of a precessing, eccentric instability in the accretion disk (Skillman et al. 1998, and references therein). In this respect, it is interesting to note the sudden appearance of an additional signal in the second part of SWASP data ($P2 = 0.137206$ d; the superhump period?) with a period close to the first period (P1 = 0.136621 d; the orbital period?), which might be interpreted in this vein. However, it is impossible to draw any definite conclusions concerning the nature of the observed period instability without further photometric and spectroscopic monitoring.

Furthermore, the observed secondary variability on a time-scale of ~26 days is in agreement with the phenomenon of stunted outbursts, which are prone to exhibiting pseudo-periods and have been observed in old novae and nova-like CVs (Honeycutt 2001). They are fairly common among VY Scl variables like e.g. FY Per and V794 Aql, which both show secondary variability on a timescale of about 20-25 days (Honeycutt 2001; Honeycutt and Kafka 2004). In this respect, it is interesting to note that – just before the onset of the fading event – the outbursts seem to occur in shortening intervals, and their amplitude increases. It is intriguing to surmise that this behaviour, which might results from mounting disk instabilities, is in some way connected to the ensuing cessation of mass transfer, which leads to the observed drop in brightness. However, an investigation of this is beyond the scope of the present paper.

Similar to the mechanism in dwarf novae, stunted outbursts are interpreted as being due to accretion disk instabilities seen against the background of a high optical brightness state which diminishes the amplitude of the outbursts (e.g. Honeycutt 2001). Generally, the observed light curve features are in good agreement with observations of other VY Scl variables (cf. Honeycutt 2001; Honeycutt and Kafka 2004).

## 4. Conclusion

Taking into account all available data, we feel confident in presenting 1SWASP J022916.91-395901.4 as a promising VY Sculptoris variable candidate. Our classification has been based mainly on

(i)    colour indices ((B-V) ≈ 0.0, (J-K) ≈ 0.7);
(ii)    the star's position in ROSAT hardness ratio space (cf. Figure 6);
(iii)    the observed, complex variability pattern that is characteristic of this type of variable star, in particular

- a. the significant fading event of ~3.5 mag (V*) between HJD 2454700 and HJD 2455200;
- b. the observed regular variability (P ≈ 0.137 days; ΔV* ≈ 0.2 mag), which is reminiscent of the orbital periods of VY Scl stars (~0.133 d < $P_{orb}$ < ~0.166 d);
- c. the observed semi-regular brightenings (P ≈ 26 days; ΔV* ≈ 1 mag), which are strongly reminiscent of the phenomenon of stunted outbursts seen frequently in VY Scl stars.

However, our conclusion needs to be verified by spectroscopic studies, which we herewith strongly encourage.

**Acknowledgements:**
This research has made use of the SIMBAD and VizieR databases operated at the Centre de Données Astronomiques (Strasbourg) in France, of the AAVSO International Variable Star Index (VSX) and of the Two Micron All Sky Survey (2MASS), which is a joint project of the University of Massachusetts and the Infrared Processing and Analysis Center/California Institute of Technology, funded by the National Aeronautics and Space Administration and the National Science Foundation. This paper has also made use of data from the DR1 of the WASP data as provided by the WASP consortium, and the computing and storage facilities at the CERIT Scientific Cloud, reg. no. CZ.1.05/3.2.00/08.0144 which is operated by Masaryk University, Czech Republic. It is a pleasure to acknowledge John Greaves and Sebastián Otero for helpful discussions. The authors thank the anonymous referees for helpful suggestions that helped to improve the paper.

Zacharias, N.; Finch, C. T.; Girard, T. M. et al., 2012, UCAC4 Catalogue, VizieR On-line Data Catalog, I/322